\documentclass[review]{elsarticle}

\usepackage{graphicx}
\usepackage{amssymb}
\usepackage{amsthm}
\usepackage{amsmath}
\usepackage{subfigure}

\newtheorem{definition}{Definition}
\newtheorem{lemma}{Lemma}
\begin{document}

\begin{frontmatter}

\title{\large{Homotopy perturbation method for fractional-order Burgers-Poisson equation}}

\author[]{Caibin Zeng \corref{cor1}}
\ead{zeng.cb@mail.scut.edu.cn}  \cortext[cor1]{Corresponding author.
Tel: +86 20 87110448; fax: +86 20 87110448.}
\author[]{Qigui Yang}
\author[]{Bengong Zhang}

\address{\footnotesize{School of Mathematical Sciences, South China
University of Technology, Guangzhou 510640,
 PR China}}

\begin{abstract}
In this paper, the fractional-order Burgers-Poisson equation is
introduced by replacing the first-order time derivative by
fractional derivative of order $\alpha$. Both exact and approximate
explicit solutions are obtained by employing homotopy perturbation
method. The numerical results reveal that the proposed method is
very effective and simple for handling fractional-order differential
equations.
\end{abstract}


\end{frontmatter}
\section{Introduction}
\label{intro} In 2004, the Burgers-Poisson (BP) equation has firstly
been proposed to describe the unidirectional propagation of long
waves in dispersive media \cite{fellnerand04}, denoted by
\begin{equation}
u_t+uu_{x}=\varphi_{x}, \label{e:1}
\end{equation}
\begin{equation}
\varphi_{xx}=\varphi+u, \label{e:2}
\end{equation}
where $\varphi$ and $u$ depend on $(t, x) \in (0,\infty) \times
{R}$, and subscripts denote partial derivatives. In order to well
study BP equation, by applying $1-\partial^2_x$ to (\ref{e:1}) and
using (\ref{e:2}) on the resulting right-hand side, we rewrite the
BP system as a single differential equation for $u$:
\begin{equation}
u_t-u_{xxt} + u_x + uu_x = 3u_xu_{xx} + uu_{xxx}.\label{e:3}
\end{equation}
Because BP equation is a shallow water equation modeling
unidirectional water wave subject to weaker dispersive effects than
the KdV equation. This means that BP equation is very important in
the field of mathematical physics. In Ref.\cite{fellnerand04}, the
authors turned out that BP equation features wave breaking in finite
time, a local existence result for smooth solutions and a global
existence result for weak entropy solutions were further proved.
Later on, BP equation was also proposed by Fetecau and Levy
\cite{fetecau05} by Pad\'{e} (2,2) approximation of the phase
velocity that arises in the linear water wave theory. In
Ref.\cite{turgay07}, the authors used classical Lie method to
construct group invariant solutions. Moreover, variational iteration
method (VIM) was applied to study the numerical solutions of BP
equation \cite{hizel07}.

Recently, fractional calculus has been extensively applied in many
fields \cite{WBG03}. Many important phenomena are well described by
fractional differential equations in electromagnetics, acoustics,
viscoelasticity, electrochemistry and material science. That is
because of the fact that, a realistic modelling of a physical
phenomenon having dependence not only at the time instant, but also
the previous time history can be successfully achieved by using
fractional calculus. In particular, Wang \cite{Wang07,Wang08}
employed the homotopy perturbation method (HPM) to solve the
classical fractional KdV and KdV-Burgers equations, respectively.
Momani \cite{Momani05} solved fractional KdV equation via the
Adomian decomposition method (ADM) and Momani \emph{et al}.
\cite{Odibat08} applied variational iteration method (VIM) to solve
the space and time fractional KdV equation. In the above methods,
HPM provides an effective procedure for explicit and numerical
solutions of a wide and general class of differential systems
representing real physical problems. Then HPM has been widely used
by other authors \cite{Momani07,Abdulaziz08} as well as their
referees to solve fractional order differential equations.

In this paper, we introduce fractional-order into the BP equation
(\ref{e:3}) by replacing the first-order time derivative by
fractional derivative of order $\alpha$, then we obtain the
fractional Burgers-Poisson (fBP) equation of the form
\begin{equation}
D_t^{\alpha}u-D_t^{\alpha}u_{xx}+u_x+u u_x-(3u_x u_{xx}+u
u_{xxx})=0, \quad t>0, 0<\alpha \le 1,\label{e:8}
\end{equation}
where $\alpha$ denotes the order of the fractional time-derivative
in the Caputo sense. The function $u(x,t)$ is assumed to be a causal
function of time and space, $i.e.$ vanishing for $t < 0$ and $x <
0$. The general response expression contains a parameter describing
the order of the fractional derivative that can be varied to obtain
various responses. In the case of $\alpha=1$, the fractional Eq.
(\ref{e:8}) reduces to the classical BP equation (\ref{e:3}).
Furthermore, HPM will be employed to obtain both exact and
approximate explicit solutions of fBP equation.

\section{Preliminaries}
\label{sec:1} We first give the definitions of fractional-order
integration and fractional-order differentiation \cite{Pod99}. For
the concept of fractional derivative, we will adopt Caputo's
definition, which is a modification of the Riemann-Liouville
definition and has the advantage of dealing properly with initial
value problems.
\begin{definition}
A real function $f(t$), $t
> 0$, is said to be in the space $C_{\mu}$, $\mu \in R$, if there
exists a real number $p > \mu$, such that $f(t)=t^p f_1(t)$, where
$f_1(t) \in C(0,\infty)$, and it is said to be in the space $C_n$ if
and only if $f^n \in C_{\mu}$, $n \in N$.
\end{definition}
\begin{definition}
The Riemann-Liouville fractional integral operator of order
$\alpha$, $J^{\alpha}$, of a function $f \in C_{\mu}$, $\mu \ge -1$,
is defined as
\begin{equation}
J^{\alpha} f(t) = \frac{1}{{\Gamma (\alpha)}}\int_0^t
{(t-\tau)}^{\alpha-1} f(\tau) d\tau, \ \ (\alpha>0, t>0).
\label{e:4}
\end{equation}
\end{definition}
\begin{definition}
The fractional derivative of $f(t)$ in Caputo's sense is defined as
\begin{equation}
D^{\alpha} f(t) = \frac{1}{{\Gamma (m - \alpha)}}\int_0^t
{\frac{{{f^{(m)}}(\tau )}}{{{{\left( {t - \tau } \right)}^{\alpha +
1 - m}}}}} d\tau , \label{e:5}
\end{equation}
where $m-1 < \alpha \le m$, $m \in N$, $t > 0$, $f \in C_{-1}^{m}$.
\end{definition}
\begin{definition}
For m to be the smallest integer that exceeds $\alpha$, the Caputo
time-fractional derivative operator of order $\alpha > 0$ is defined
as
\begin{equation}
D_t^{\alpha} u(x,t) = \left\{%
 \begin{array}{ll}
 \frac{1}{\Gamma(m-\alpha)} \int_{0}^{t}{(t-\tau)}^{m-\alpha-1} \frac{\partial^{m}u(x,\tau)}{\partial \tau^{m}} d\tau, &\textrm{if } m-1< \alpha <m,\\
 \frac{\partial^{m}u(x,t)}{\partial t^{m}},  &\textrm{if }\alpha = m \in
 N.
 \end{array}%
 \right.\label{e:6}
\end{equation}
\end{definition}

\begin{lemma}\label{lemma:1}
If $m-1 < \alpha \leq m$, $m \in N$, and $f \in C_{\mu}^{m}$, $\mu
\ge -1$, then
\begin{equation}
J^{\alpha}D^{\alpha} f(x)= f(x)- \sum\limits_{k=0}^{m-1}
f(0^{+})\frac{x^k}{k!}.\label{e:7}
\end{equation}
\end{lemma}

\section{Application of HPM to fBP equation}
\label{sec:2} Homotopy perturbation method (HPM) is first proposed
by He \cite{he97}. It is a powerful mathematic tool to solve
nonlinear problems, especially engineering problems. To further
complement HPM, He has developed it \cite{he03,he08} recently.

In what follow, the HPM is used to study the fBP equation
(\ref{e:8}) with the initial condition
\begin{equation}
u(x,0)=x.
\end{equation}
The exact solution of BP equation (\ref{e:3}), the special case of
fBP equation (\ref{e:8}) when $\alpha=1$, is given by classical Lie
method (see Ref.\cite{turgay07})
\begin{equation}
u(x,t)=\frac{1+x}{1+t}-1.\label{e:10}
\end{equation}
In view of HPM, the homotopy is constructed as following
\begin{equation}
(1-p)D_t^{\alpha}u+p(D_t^{\alpha}u-D_t^{\alpha}u_{xx}+u_x+u
u_x-(3u_x u_{xx}+u u_{xxx}))=0,\label{e:11}
\end{equation}
or
\begin{equation}
D_t^{\alpha}u+p(-D_t^{\alpha}u_{xx}+u_x+u u_x-(3u_x u_{xx}+u
u_{xxx}))=0,\label{e:12}
\end{equation}
where $p \in [0,1]$ is an embedding parameter. By utilizing the
parameter $p$, the solution $u(x,t)$ can expanded in the following
form
\begin{equation}
u(x,t)=u_0(x,t)+pu_1(x,t)+p^2 u_2(x,t)+p^3
u_3(x,t)+\cdots.\label{e:13}
\end{equation}
Setting $p=1$ gives the approximate solution
\begin{equation}
u(x,t)=u_0(x,t)+u_1(x,t)+ u_2(x,t)+u_3(x,t)+\cdots.\label{e:14}
\end{equation}
Now, substituting (\ref{e:14}) into (\ref{e:12}), and equating the
terms with the identical powers of $p$, gives
\begin{equation}
 \begin{array}{l}
 p^0 : D_t^{\alpha} u_0 = 0 , \ \ u_0(x,0)=x;  \\
 p^1 : D_t^{\alpha} u_1 - D_t^{\alpha} u_{0xx}+u_{0x}+ u_0 u_{0x}-3
u_{0x}u_{0xx}-u_0 u_{0xxx}= 0 , \ \ u_1(x,0)=0;  \\
 p^2 : D_t^{\alpha} u_2 - D_t^{\alpha} u_{1xx}+u_{1x}+ (u_0
u_{1x}+u_1 u_{0x})-3 (u_{0x}u_{1xx}+u_{1x}u_{0xx})\\ \qquad -(u_0
u_{1xxx}+u_1 u_{0xxx})= 0 , \ \ u_2(x,0)=0; \\
p^3 : D_t^\alpha  u_3 - D_t^{\alpha} u_{2xx}+u_{2x}+ (u_0 u_{2x}+u_1
u_{1x}+u_2 u_{0x})\\
\qquad -3 (u_{0x}u_{2xx} +u_{1x}u_{1xx}+u_{2x}u_{0xx})-(u_0
u_{2xxx}+u_1 u_{1xxx}+u_2 u_{0xxx})= 0 , \\
\qquad u_3(x,0)=0;\\
\vdots
 \end{array} \label{e:15}
 \end{equation}
Applying the operator $J^\alpha$ on both sides of equations in
(\ref{e:15}) and utilizing Lemma \ref{lemma:1} , yields
\begin{equation}
\begin{array}{l}
u_0 (x,t) = x;\\
u_1 (x,t) = - \frac{1+x}{\Gamma(1+\alpha)}t^\alpha;\\
u_2 (x,t) = \frac{2(1+x)}{\Gamma(2+\alpha)}t^{2 \alpha}; \\
u_3 (x,t) = -\frac{(1+x)(4
\Gamma(1+\alpha)^2+\Gamma(2+\alpha))}{\Gamma(1+\alpha)^2
\Gamma(3+\alpha)}t^{3 \alpha}\\
\vdots
\end{array} \label{e:16}
\end{equation}
Since
\begin{equation}
\begin{array}{l}
D_t^\alpha  u_j - D_t^{\alpha} u_{(j-1)xx}+u_{(j-1)x}+
\sum\limits_{i=0}^{j-1}u_i u_{(j-i-1)x}
-3\sum\limits_{i=0}^{j-1}u_{ix}u_{(j-i-1)xx}
\\
\qquad -\sum\limits_{i=0}^{j-1}u_i u_{(j-i-1)xxx}= 0 \label{e:17}
\end{array}
\end{equation} and
$u_j(x,0)=0,$ the other arbitrary $u_j(x, t)$ $(j \ge 4)$ can be
calculated in the same manner by symbolic software programme
\emph{Mathematica}. If only the first four approximations of
equation (\ref{e:14}) are sufficient. Then the approximate explicit
solution of equation (\ref{e:8}) will be expressed as
\begin{equation}
u(x,t)=x-\frac{1+x}{\Gamma(1+\alpha)}t^\alpha+\frac{2(1+x)}{\Gamma(2+\alpha)}t^{2
\alpha}-\frac{(1+x)(4
\Gamma(1+\alpha)^2+\Gamma(2+\alpha))}{\Gamma(1+\alpha)^2
\Gamma(3+\alpha)}t^{3 \alpha}.\label{e:18}
\end{equation}

In the other word, the exact solution for fBP equation when
$\alpha=1$ can be obtained by HPM. In face, if $\alpha=1$, equation
(\ref{e:16}) will read
\begin{equation}
\begin{array}{l}
u_0 (x,t) = x;\\
u_1 (x,t) = -(1+x)t;\\
u_2 (x,t) = (1+x)t^2; \\
u_3 (x,t) = -(1+x)t^3.
\end{array} \label{e:19}
\end{equation}
According to (\ref{e:17}), $u_j$ satisfies
\begin{equation}
u_j (x,t) = (-1)^{j}(1+x)t^j.
\end{equation}
Then the exact solution will be expressed as
\begin{equation}
u(x,t) = \lim_{n\rightarrow \infty} \sum\limits_{j=0}^{n}u_{j}(x,t)=
\frac{1+x}{1+t}-1, \label{e:20}
\end{equation}
which is just the same as for the classical Lie method
\cite{turgay07} and the VIM \cite{hizel07}.

\begin{table}[htb]\renewcommand{\arraystretch}{1} \addtolength{\tabcolsep}{15pt}
\caption{Comparison with the exact solution and the four-term
approximation solution.}
\begin{center}
\begin{tabular}{@{\extracolsep{\fill}}llll}
\hline
{$(x,t)$} & {$u_{exact}$} & {$u_{HPM}$} & {$|u_{exact}-u_{HPM}|$}\\
\hline
{(2,0.3)}     &  {0.528462}  & {0.526}  & {0.002462}\\
{(2,0.35)}  &  {0.481481}  & {0.45925}  & {0.022231}\\
{(2,0.4)}  &  {0.428571}   & {0.392} & {0.036571}\\
{(2,0.45)}  &  {0.37931}   & {0.32275} & {0.05656}\\
{(2,0.5)}       &  {0.333333}  & {0.25}  & {0.083333}\\
{(0.9,0.2)}       &  {0.583333}  & {0.5804}  & {0.002933}\\
{(1.2,0.2)}       &  {0.833333}  & {0.8304}  & {0.002933}\\
{(1.5,0.2)}       &  {1.08333}  & {1.08}  & {0.00333}\\
{(1.8,0.2)}       &  {1.33333}  & {1.3296}  & {0.00373}\\
{(2,0.2)}       &  {1.5}  & {1.496}  & {0.004}\\
\hline
\end{tabular}
\end{center}
\end{table}

\section{Numerical simulation}
\label{sec:3} In order to illustrate the approximate solution is
efficiency and accuracy, we will give explicit value of the
parameters $x, t$. Then calculate the two solutions and make a
comparison between them. Taking $x=2$ for different value of $t$ and
$t=0.2$ for different value of $x$, we calculate the numerical
solutions of the two solutions which are given in (\ref{e:10}) and
(\ref{e:18}). We list their numerical solutions in Table 1.

From the numerical solutions in the Table 1, it can be seen that at
the same time $x$, the value of the approximates solutions and the
exact solutions are quite close. Also, when the value of $t$
decreases the approximate solutions are more and more closed to the
exact solutions. This shows the approximate solution is efficiency.
It is also suggested that HPM is a powerful method for solving
fractional differential equation with fully nonlinear dispersion
terms.

\section{Conclusions}
\label{sec:4} In this paper, both exact and approximate explicit
solutions of fractional-order Burgers-Poisson equation are obtained
by employing homotopy perturbation method. The comparison of
numerical solution and exact solution demonstrates that the proposed
method is very effective and simple for solving solutions of
fractional differential equations. It should be pointed out that
detailed studies of fractional-order Burgers-Poisson equation are
only beginning. The periodical waves, peakons, fractional Hamilton
structure and other properties are still open. We hope that this
work is a step in this direction and some traditional analytic
method for nonlinear differential equations of integer order can be
extended to fractional-order equations.

\section*{Acknowledgements}

This work was supported by the National Natural Science Foundation
of China (No. 10871074).

\end{document}